\documentclass{aa}  

\usepackage{graphicx}
\usepackage{txfonts}
\def\simless{\mathbin{\lower 3pt\hbox
     {$\rlap{\raise 5pt\hbox{$\char'074$}}\mathchar"7218$}}}   %< or of order
\def\simmore{\mathbin{\lower 3pt\hbox
     {$\rlap{\raise 5pt\hbox{$\char'076$}}\mathchar"7218$}}}   %> or of order

\def\msun{~{\rm M}_\odot}

\def\ergs{erg s$^{-1}$}
\def\ergcs{erg cm$^{-2}$ s$^{-1}$}

\newcommand{\vtau}  {1A\,0535+262}

\begin{document}

\title{X-ray spectral-timing variability of 1A\,0535+262 during the 2020 giant outburst}

\subtitle{}

\author{P. Reig\inst{1,2}
\and
R.C. Ma\inst{3,4}
\and
L. Tao\inst{3}
\and
S. Zhang\inst{3}
\and
S. N. Zhang\inst{3,4}
\and
V. Doroshenko\inst{5,6}
}

\institute{Institute of Astrophysics, Foundation for Research and Technology-Hellas, 71110 Heraklion, Crete, Greece\\
\email{pau@physics.uoc.gr}          
\and
University of Crete, Physics Department, 70013 Heraklion, Crete, Greece
\and
Key Laboratory of Particle Astrophysics, IHEP, Chinese Academy of Science,
Beijing 10049, China 
\and
University of Chinese Academy of Sciences, Chinese Academy of Sciences, Beijing 100049, 
People's Republic of China 
\and
Institut f\"ur Astronomie und Astrophysik, Kepler Center for Astro and Particle
Physics, Eberhard Karls, Universitat, 
Sand 1, D-72076 T\"ubingen, Germany
\and
Space Research Institute of the Russian Academy of Sciences, Profsoyuznaya Str. 84/32, Moscow 117997, Russia
}

\date{}

% \abstract{}{}{}{}{} 
% 5 {} token are mandatory
 
\abstract
{The Be/X-ray binary 1A\,0535+262 underwent a giant X-ray outburst
in November 2020, peaking at $\sim1\times 10^{38}$ erg s$^{-1}$ (1--100
keV, 1.8 kpc), the brightest outburst recorded for this source so far. The source 
was monitored over two orders of magnitude in luminosity with {\it Insight}-HXMT,
which allowed us to probe the X-ray variability in an unprecented range of
accretion rates.  }
{Our goal is to search for patterns of correlated spectral and timing behavior
that can be used to characterize the accretion states in hard X-ray transient
pulsars.}
{We have studied the evolution of the spectral continuum emission using 
hardness-intensity diagrams and the aperiodic variability of the source by analyzing power
density spectra. We have used phenomenological models to fit the various
broad-band noise components. } 
{The hardness-intensity diagram displays three distinct branches that can be identified
with different accretion regimes. The characteristic frequency of the noise
components  correlates with the luminosity. Our observations cover the highest
end of this correlation, at luminosities not previously sampled. We have found
evidence for a flattening of the correlation at those high  luminosities, which
might indicate that the accretion disk reached the closest distance from the
neutron star surface during the peak of the outburst. We also find evidence for 
hysteresis in the spectral and timing parameters: at the same luminosity level,
the spectrum is harder and the characteristic noise frequency larger during the  
rise than during the decay of the outburst.   }
{
As in black hole binaries and low-mass X-ray binaries, the hardness-intensity
diagram represents a useful diagnostic tool to define the source state in an
accreting pulsar. Our timing analysis confirms previous findings from spectral
analysis of a hysteresis pattern of variability where the spectral and timing
parameters adopt different values for similar luminosity depending on whether
the source is on the rising or decaying phase of the outburst.
}

\keywords{accretion, accretion disks -- 
X-ray binaries: neutron stars -- 
Be stars -- 
X-ray spectra 
}

\authorrunning{Reig P.}

\titlerunning{X-ray spectral-timing variability of 1A\,0535+262 }

   \maketitle
%
%-------------------------------------------------------------------

 %-----------------------------------
\begin{table*}
\caption{Log of the spectroscopic observations.}
\label{log}
\begin{center}
\begin{tabular}{lcclccc}
\noalign{\smallskip}	\hline \noalign{\smallskip}
Observation	&Num. of&MJD  		&Date	&\multicolumn{3}{c}{On-source time (s)}	\\
		&exposures	&(start)	&	&$LE$	&$ME$	&$HE$		 \\
\noalign{\smallskip}\hline\noalign{\smallskip}
P0304099001  	&1	&59159.109	&2020-11-06	&2741   &4590	&4851		  \\  
P0304099002	&2	&59161.097	&2020-11-08	&3253   &6120	&1354		     \\  
%P030409900201  		&59161.0971   &2020-11-08   &2653.0    &4890.0	  &   0.0   \\  
%P030409900202  		&59161.2696   &2020-11-08   & 600.0    &1230.0	  &1354.0   \\  
P0304099003	&2	&59163.019	&2020-11-10	&3153	&7170	&5852		   \\
%P030409900301  		&59163.0192   &2020-11-10   &2553.0    &4276.4	  &4192.5   \\  
%P030409900302  		&59163.1909   &2020-11-10   & 660.0    &1620.0	  &1660.0   \\  
P0304099004  	&1	&59165.074   &2020-11-12   &2775    &4320  &1725.5		   \\  
P0304099005  	&1	&59167.128   &2020-11-14   &1527    &2850  &2848.5		     \\  
P0304099006  	&1	&59169.050   &2020-11-16   &2702    &3360  &3542.0		     \\  
P0304099007  	&1	&59171.104   &2020-11-18   & 935    &1290  &1412.0		     \\  
P0304099008  	&1	&59174.019   &2020-11-21   & 899    &2100  &2171.0		     \\  
P0304099009  	&1	&59177.066   &2020-11-21   &	0    & 360  & 504.0		     \\  
P0304099010  	&1	&59180.048   &2020-11-21   &	0    & 600  & 577.5		     \\  
P0304099011  	&1	&59183.097   &2020-11-30   &1260    &2640  &1614.5		     \\  
P0304099012  	&1	&59186.015   &2020-12-03   &1723    &3420  &1157.5		  \\   
P0304099013  	&1	&59189.069   &2020-12-06   &3675    &6720  &5454.0		  \\   
P0304099014  	&1	&59191.991   &2020-12-08   &2280    &6000  &3709.0		  \\   
P0304099015  	&1	&59195.104   &2020-12-12   &2292    &2910  &1381.0		  \\   
P0304099016  	&1	&59198.018   &2020-12-15   &1080    &3000  &2160.0		  \\   
P0304099017  	&1	&59201.065   &2020-12-18   &1597    &2160  &  29.0		  \\   
P0304099018  	&1	&59204.046   &2020-12-21   &1717    &2250  &   0.0		  \\   
\noalign{\smallskip}	\hline						   
P0314316001   &13     &59167.327    &2020-11-14   &20145      &37267  &24367		 \\
P0314316002   &14     &59169.249    &2020-11-15   &21063      &37650  &28733		 \\
P0314316003   &20     &59171.303    &2020-11-18   &27612      &51930  &37695		 \\
P0314316004   &21     &59174.218    &2020-11-21   &18535      &52920  &42073		 \\
P0314316005   &21     &59177.265    &2020-11-22   &25343      &52380  &41156		\\
P0314316006   &21     &59180.247    &2020-11-27   &34859      &51150  &40294		 \\
P0314316008   &20     &59183.296    &2020-11-30   &40834      &57828  &46878		 \\
P0314316009   &14     &59186.214    &2020-12-03   &45422      &77842  &50860		 \\
P0314316010   &16     &59189.269    &2020-12-06   &31000      &70950  &54650		 \\
P0314316011   &22     &59192.190    &2020-12-09   &34927      &58830  &42155		 \\
P0314316012   &20     &59195.303    &2020-12-12   &30240      &51120  &34852		 \\
P0314316013   &21     &59198.217    &2020-12-15   &29576      &51240  &37231		 \\
P0314316014   &13     &59201.264    &2020-12-18   &25669      &50940  &37196		 \\
\noalign{\smallskip}	\hline						   
\end{tabular}								   
\end{center}								   
\end{table*}								   
%----------------------------------------				   

\section{Introduction}

\vtau\ was one of the first X-ray pulsars to be discovered. The first
observations date back to 13 April 1975 with the {\it Ariel V} mission
\citep{rosenberg75,coe75}.  In these first observations, the source was
identified as an X-ray pulsar with a spin period of 104 s. Based on the
positional coincidence  with the best X-ray position, several authors
\citep{hudec75,murdin75} noted the V=9 mag star V725 Tau/HD245770 as a possible
optical counterpart.  The first suggestion of a binary system was given by
\citet{rappaport76} by studying the variation in the 104-s periodicity. Their
results were consistent with a neutron star and a OB star companion.
\citet{janot87} derived a spectral type B0IIIe. Because of its high X-ray
variability and optical brightness, \vtau\ is one of the best studied Be /X-ray
binaries (BeXB) with observations across the entire electromagnetic spectrum. 

The source is not detected neither in the radio band \citep{tudose10,migliari11}
nor in the $\gamma$-ray band above $E>0.1$ GeV \citep{acciari11}.
Near-IR spectroscopy of the object shows that the JHK spectra are dominated by
the emission lines of hydrogen Brackett and Paschen series and HeI lines at
1.0830, 1.7002 and 2.0585 $\mu$m. Infrared excess is observed and attributed to
the circumstellar disk around the Be star. The amplitudes of the JHK bands
variations are about 0.1 mag on time scales of years
\citep{persi79,gnedin83,clark98b,haigh99,haigh04,naik12,taranova17}.

The UV observations served to estimate a mass loss rate through stellar wind from
the early-type companion of $\sim 10^{-8}$ $\msun$ yr$^{-1}$ and an effective
temperature of 26000 K. The depth of the 2200 \AA\ interstellar extinction
feature gave a color excess of $E(B-V)=0.72$
\citep{wu83,deLoore84,payne87,clark98a}.

\vtau\ displays long-term optical photometric
\citep{hao96,clark99,lyuty00,zaitseva05} and spectroscopic \citep{yan12}
variability, possibly associated with mass ejection episodes as well as cyclic
spectroscopic variability in the H$\alpha$ and other lines, which is interpreted
as global one-armed oscillation in the disk \citep{clark98a,camero12}.
Asymmetric spectral lines has been associated with warped circumstellar disks
during giant X-ray outbursts \citep{moritani13}.

In the X-ray band, \vtau\ is one of the most active BeXB
with frequent X-ray outbursts. \vtau\ exhibits the two types of outbursts known
to BeXBs \citep{motch91}. Regular type I outbursts show a moderate increase in
X-ray flux ($L_X \simless 10^{37}$ \ergs), occur near periastron passage, and
last for a fraction of the orbit. Giant or type II outbursts are significantly
brighter ($L_X \simmore 10^{37}$ \ergs), do not occur at any preferential
orbital phase and may last for several orbits. The November 2020 event is the
fourth major outburst in the past 16 years and the brightest ever recorded.

Accreting X-ray pulsars exhibit strong X-ray variability. Periodic variability
is related to the rotation of the neutron star and manifests as pulsations of
the order of seconds. Another example of (Quasi) periodic variability is the type
I X-ray outburst, which are orbitally modulated. The slowest time scales are
linked to the mass transfer process between the Be star and the neutron star and
manifest as unpredictable giant (or type II) X-ray outbursts on time scales of
years. Time scales attributed to the accretion process vary in the range from
fraction of a second to hours and manifest as broad-band noise in the power
spectrum \citep{revnivtsev09,mushtukov19}. In addition, quasi-periodic
oscillations (QPO) in the millihertz range are commonly detected
\citep{james10}. In this work we have studied the spectral changes and  the 
variability of the broad-band noise of \vtau\ during the November 2020 X-ray
outburst.

%---------------------------FIG 1-------------------------------------- 
\begin{figure*}
\centering
\includegraphics[width=14cm]{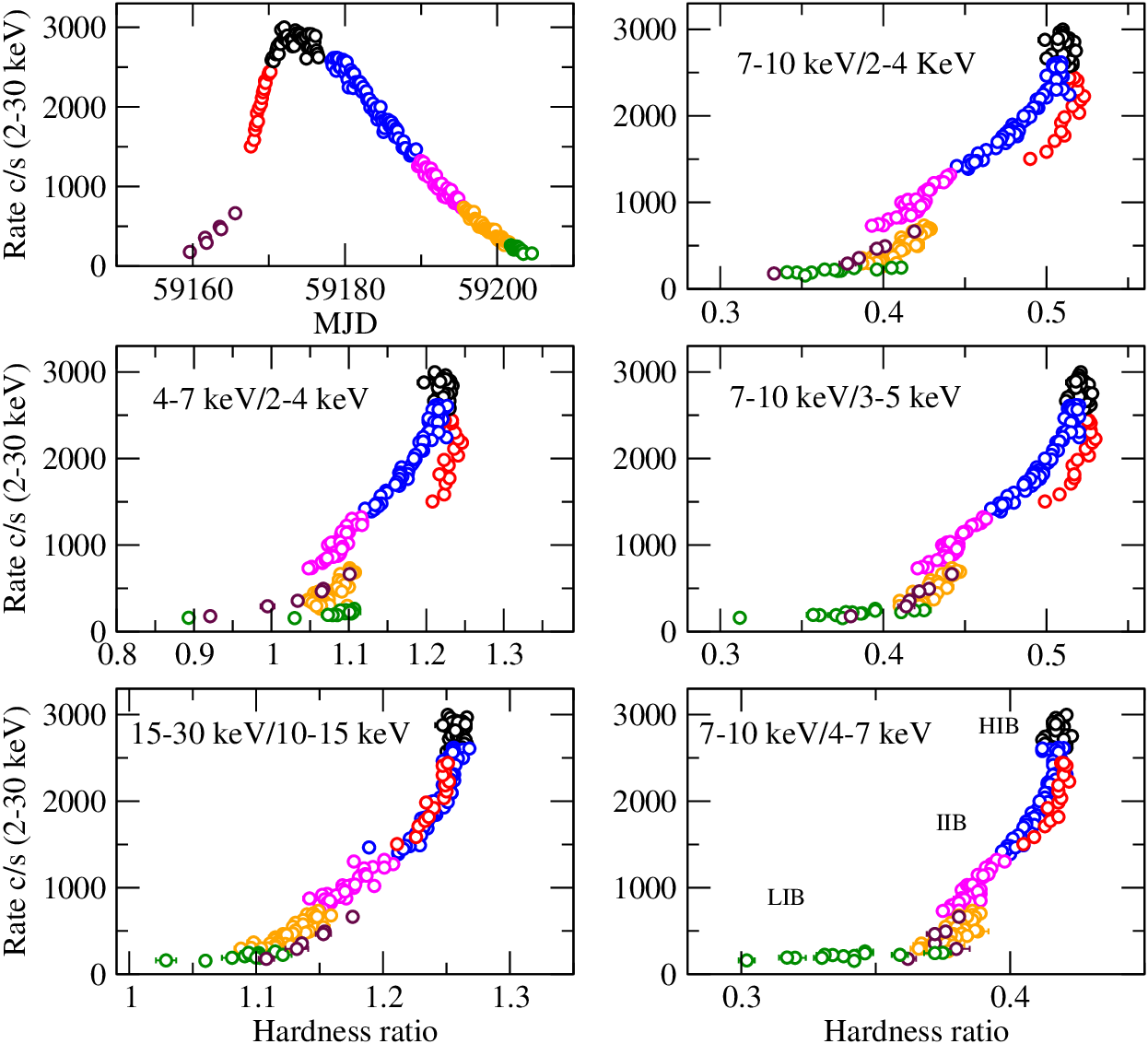}
\caption{Outburst light curve and hardness-intensity diagrams for
various selection of the hardness ratio, shown on the top left corner of each
diagram. Different colors represent different instances of the outburst as shown
in the light curve (top left panel).}
\label{hid}
\end{figure*}
%-----------------------------------------------------------------

\section{Observations}

The source was observed  by the Hard X-ray Modulation Telescope ({\it
Insight}-HXMT, henceforth) covering the period from 6 November 2020 to  24
December 2020. {\it Insight}-HXMT was  launched on 15th June 2017 from JiuQuan, China,
runs in a low earth orbit with an altitude of 550 km and an inclination angle of
43$^{\circ}$ \citep{zhang20}. It carries three instruments on board: the High
Energy X-ray telescope (HE) uses 18 NaI(Tl)/CsI(Na) scintillation detectors and
it is sensitive to X-rays in the 20-250 keV band. It has a total geometrical
area of about 5100 cm$^2$  and the energy resolution is ~15\% at 60 keV
\citep{liu20}; the Medium Energy Xray telescope (ME) consists of 1728 SiPIN
detectors to detect photons in the 5-30 keV band using a total geometrical area
of 952 cm$^2$ \citep{cao20}; the Low Energy Xray detector (LE) contains 96 SCD
detectors suitable for photons with energies in the range 1-15 keV and a
geometrical area of 384 cm$^2$ \citep{chen20}. The three payloads are integrated
on the same supporting structure to achieve the same pointing direction, thus
they can simultaneously observe the same source.

We analyzed observations from two different proposals: P0304099 (PI: P. Reig)
and P0314316 (PI: Core Science Team). P0304099 consisted of 18 snapshots from 6
November 2020 to  21 December 2020. The observations were made every 2 days 
during the rise (MJD 59159--59171) and every 3 days during the decay (MJD
59171.1--59204.0). Owing to a very high count rate, no event file was generated
for the $LE$ instrument during the exposures P0304099009 and P0304099010. Each
observation had a total duration of $\sim$10 ks, although the on-source time was
only a fraction of it. P0314316 covered the interval MJD 59167--59205. The
observations contained multiple exposures resulting in long observing intervals.
Table~\ref{log} shows the log of the HXMT observations.

The data were screened using good time intervals created with the following
criteria: the Earth elevation angle greater than 10 degrees, the cutoff
rigidity  greater than 8 GeV, the offset angle from the pointing source less
than 0.04 degrees. We also exclude the photons collected 300 s before enter and
after exit the South Atlantic Anomaly.

\section{Results}

In this work, we focus on the general shape of the X-ray spectral
continuum and the broad-band noise and their variation with X-ray luminosity. A
detailed X-ray spectral analysis, including the study of the cyclotron line was
performed by \citet{kong21}. For a detailed analysis of the milliherzt
quasi-periodic oscillation (QPO), see Ma et al. 2022 (in preparation).

\subsection{Spectral analysis}

We obtained energy spectra for each observation and each instrument. The spectra
were extracted with the sole purpose to compute the X-ray flux. For this reason,
we used a simple phenomenological model and fitted the spectra separately for
each instrument. The continuum was fitted with a power law, modified at high
energies but an exponential cutoff and at low energies by photoelectric
interstellar absorption; in addition, a discrete component corresponding to the
fluorescence line of iron at 6.4 keV was included in the LE spectral fit and a
cyclotron line at $\sim$45 keV in the HE spectral fit. These components were
model with Gaussian functions in emission  and absorption, respectively.  To
compute the X-ray luminosity, we assumed a distance to the source of 1.8 kpc
\citep{bailer-jones21}.

\subsection{Hardness/color analysis}
\label{hardness}

An X-ray color or hardness ratio is the ratio of  the photon counts between two
broad bands. It provides a model independent way to study the spectral changes
without the need to consider complex fitting procedures. On the opposite side,
it depends on the detector used, hence it is instrument dependent.

To avoid possible effects of the interstellar absorption we chose energies above
2 keV. We also avoided bands containing spectral lines, e.g 6.4 keV iron.
However, for the sake of comparison with previous studies we also included the
4--7 keV band.  Figure~\ref{hid} shows the outburst light curve (top left panel)
and the hardness-intensity diagram (HID) for various hardness ratios. The rise of the
outburst was significantly faster than the decay and it is less well sampled.
The source took 10--15 days to reach maximum flux, but then it took about a
month to go from peak to a similar flux to the first  observation. To be able to
follow the motion of the source in the HID, we distingished between the rise and
the decay of the outburst and used different colors to represent various
intervals that differ in the count rate. The following features may be noticed
when inspecting the HID:

\begin{itemize}

\item Three distinct branches can be seen in the HID.  At low count rate, the
source moves horizontally in the HID, that is, significant spectral changes
are seen for a very low change in count rate (green circles). At certain
intensity, the source turns up and moves diagonally. As the intensity increases,
the spectrum becomes harder. This branch contains most of the observations
during the rise and decay of the outburst (brown, red, blue, magenta, and
orange circles). The positive intensity-hardness ratio correlation stops near the peak
of the outburst, when the source stabilizes or even moves vertically (black
circles). 

%We shall refer to these three branches as the low-intensity branch (LIB), the
%intermediate-intensity branch (IIB), and the high-intensity branch (HIB).

\item The soft part of the spectrum, $E\simless 5$ keV, follows different
tracks during the rise and decay of the outbursts. During the rise, the spectrum
is harder than during the decay (compare red and blue points in Fig.~\ref{hid}).
This hysteresis effect decreases as the energy of the bands considered
increases. 

\item We observed a  break in hardness during the decay. At $\sim 700$ c/s ($L_x
\approx 1.7\times 10^{37}$ \ergs in the 2-30 keV energy range), the source jumps
from the decaying track to the rising track (note the discontinuity between the
magenta and the orange circles in Fig.~\ref{hid}). Again this effect is observed
only at low energy.

\end{itemize}

\subsection{Broad-band noise}

The fast temporal variability of the source is characterized by periodic
(pulsations) and aperiodic (red noise) components. To investigate the broad-band
noise components associated with the aperiodic variability, we obtained the 
power spectral density (PSD) or simply power spectrum by Fourier transform of
the light curves in the 30--100 keV band. We use the HE instrument because it
offers the largest effective area of the three instruments.  To decrease the
error in each frequency bin, the light curves were divided into segments and a
Fast Fourier Transform was computed for each segment \citep[see e.g.][]
{vanderklis89,vanderklis06}. The final power spectrum is the average of the
power spectra obtained for each segment.

The frequency interval covered by the power spectrum is given by $[1/T,\nu_{\rm
nyq}]$, where $T$ is the duration of the segment and $\nu_{\rm nyq}=1/(2\delta
t)$ is the Nyquist frequency. $\delta t$ is the time resolution of the light
curve. Because the variability of accreting X-ray pulsars at high frequencies is
strongly suppressed \citep{reig13}, most likely due to the viscous diffusion
associated to accretion rate fluctuations \citep{mushtukov19}, the time
resolution need not be too small. We set the highest frequency at 16 Hz ($\delta
t=0.03125$ s).  The length of each interval was chosen to be 256 s, hence the
minimum frequency is $\sim 0.004$ Hz.   Since our goal is to study the correlated
spectral-timing behavior of the source, we obtained PSD covering the same
intervals as in the color analysis. At least two PSD were obtained for each
interval, except for the the observations with the lowest intensity (green
circles) that given the low S/N, we generated one average PSD.

To study the broad-band noise, we removed the contribution of the pulse flux
from the light curves prior to the determination of the PSD \citep{finger96,
revnivtsev09}. For each 1024-s segment ($\sim$ 10 spin cycles), we obtained an
average pulse profile. We replicated this profile and created a light curve with
the same duration and binning as the original light curve. Then we subtracted
the replicated light curve from the original one. The removal of the pulse peak
and its harmonics were not always satisfactory, leaving some residuals. Since we
use average light curves in the 30-100 keV range,  the variability of the pulse
profiles with energy \citep{mandal20} may contribute to those residuals. 

We tried a second method consisting of simply removing the frequency bins in the
PSD most affected by the pulsations and its harmonics. The results in terms of
the value of the best-fit parameters and its dependence with luminosity were
consistent with the pulse removal methodology. 

To fit the PSD, we followed two approaches. The first approach is that generally
used in black-hole binaries. In these sources, it is common practice to describe
the timing features with a function consisting multiple Lorentzians, $L_i$,
where $i$ determines the number of the component \citep{belloni02}. In \vtau,
the {\it Insight}-HXMT PSD continuum in the 0.004--16 Hz frequency range is well
described by three zero-centered Lorentzians, while another narrow Lorentzian is
needed to fit the QPO.  The characteristic frequency of $L_i$ is denoted
$\nu_{L_i}$. This is the frequency where the component contributes most of its
variance per logarithmic frequency interval and is defined as
$\nu_L~=~\sqrt{(\nu_0^2+(FWHM/2)^2)}$, where $\nu_0$ is the centroid frequency
and FWHM is the Lorentzian full-width at half maximum. For a zero-centered
Lorentzian,  the characteristic frequency is simply half of its width.

The second approach is to use a broken power law model.
Because many accreting pulsars displays breaks in their power spectra
\citep{revnivtsev09,mushtukov19}, the broken or double broken power law model
has been used successfully in broad-band noise analysis
\citep{doroshenko14,doroshenko20}. The model parameters of a
broken power law are, in addition to the normalization, the break frequency,
$\nu_{\rm break}$  and two power low indices, $\Gamma_1$, $\Gamma_2$. A
Lorentzian is added to account for the QPO. 

\begin{figure}
\centering
\includegraphics[width=8cm]{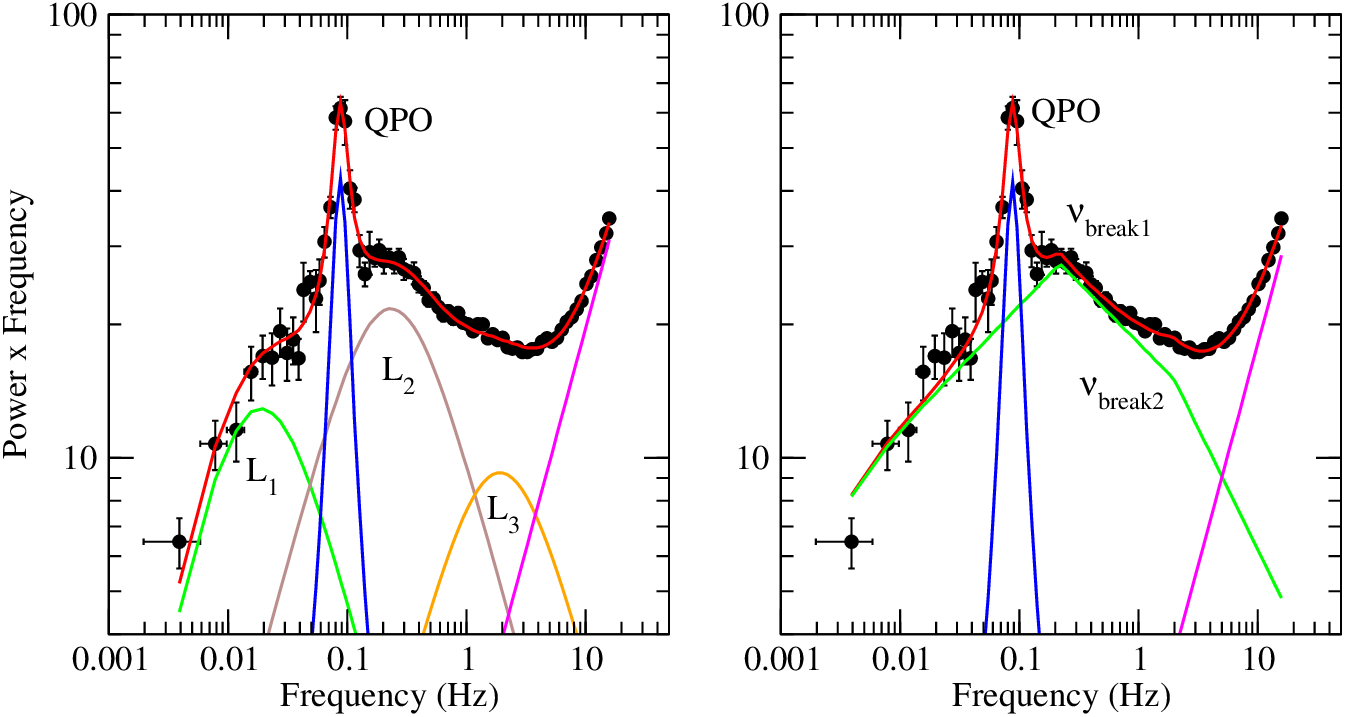}
\caption{Representative example of a power spectrum and model components used to
fit the broad-band noise: multi-Lorentian model (left) and broken power-law
model (right). The Poisson noise was fitted with a power
law with index fixed to zero (magenta line). The data correspond to observations taken during
22-25 November 2020. }
\label{psdfitmodel}
\end{figure}

The broken power-law fitted well the PSD of \vtau\ at low luminosity (count
rate below $\sim 275$ cs$^{-1}$ in 2--30 keV). Above that value or
equivalently $2\times 10^{-8}$ \ergcs\ (1--100 keV), the overall rms increases
and extra component is required. We tried fitting a double broken-power law
model, which included a second break frequency and a third power-law index. The
fit improved considerably with the reduced $\chi^2$ changing from 2--4 to 1--2
for 65 and 62 degrees of freedom, typically. The physical meaning of the higher
frequency break is not clear. It has also been observed in the super-luminous
X-ray pulsar Swift J0243.6+6124, albeit at  luminosities above the Eddington
limit \citet{doroshenko20}. Its frequency ($\sim 5-7$ Hz) is somehow higher than
the frequency observed in \vtau\ (1--3 Hz). However, this can be understood by
the natural shift of all the frequencies that characterize the aperiodic
variability of pulsars with flux and by the different magnetic field strength. 
Figure~\ref{psdfitmodel} shows a representative PSD and the components of the
two models considered. The Poisson noise was account for with a power law with
index fixed to zero. The two models give comprable results in terms of the
quality of the fit (reduced $\chi^2$), although the fitting parameters returned
by the multi-Lorentzian model presented slightly less dispersion, especially at
low luminosity. The reason may be the number of parameters involved in the fit
(nine in the multi-Lorentzian model and six in the double-broken power-law
model, excluding the QPO).

\begin{figure}
\centering
\includegraphics[width=8cm]{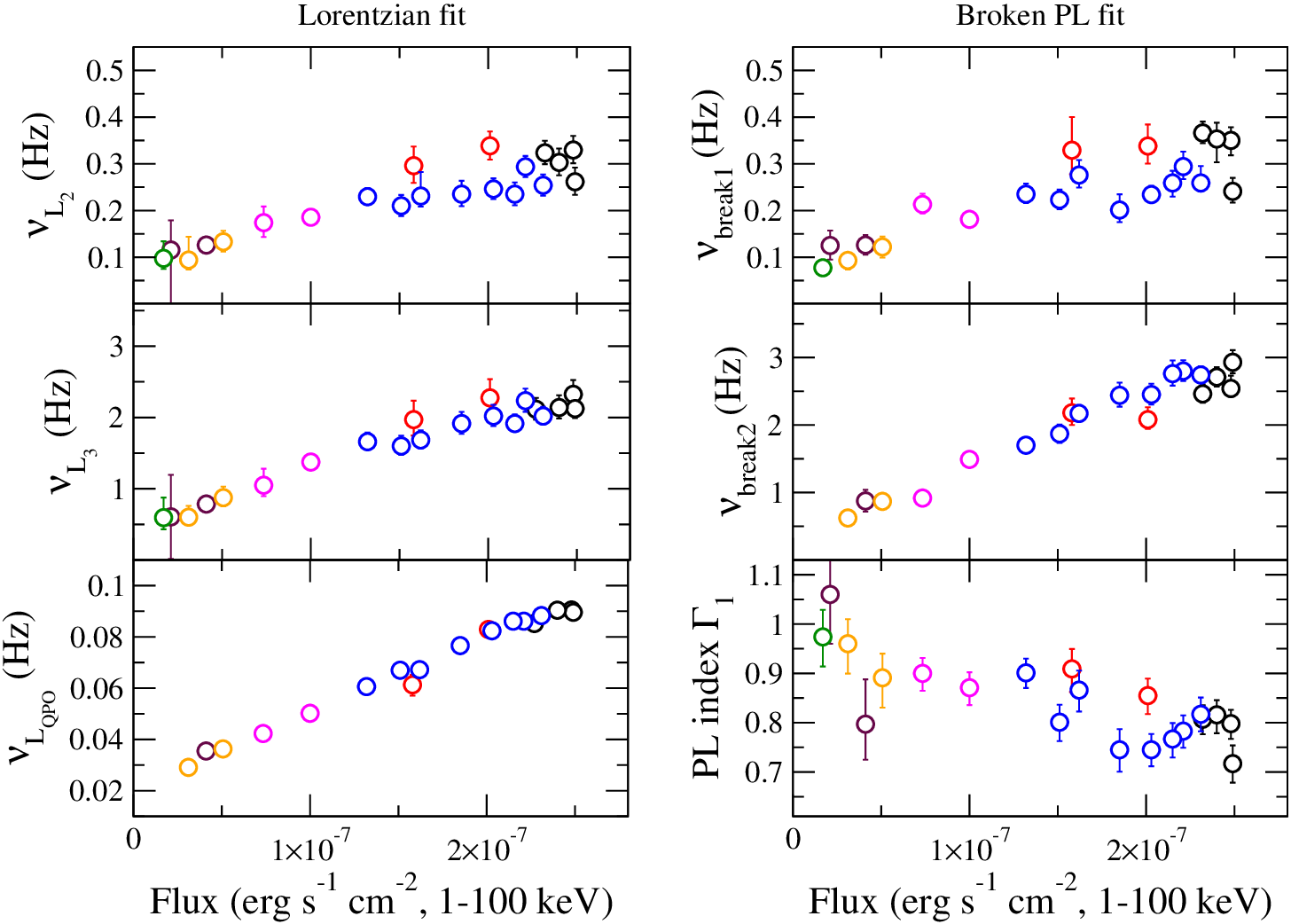}
\caption{Best-fit frequencies using the multi-Lorentzian model (left) and the
broken power-law model (right) as a function of X-ray flux. The color code as in
Fig.~\ref{hid}.}
\label{freq}
\end{figure}

Figure~\ref{freq} shows the dependence of the characteristic frequencies with
1-100 keV flux. The left panel displays the characteristic frequencies of the
multi-Lorentzian fit, while the right upper panel shows the break frequencies of
the (double) broken power law model and the QPO.  The color code is the same as
in Fig.~\ref{hid}: brown and red circles represent observations during the rise
of the outburst,  black points correspond to its peak, and the blue, magenta,
orange, and green circles to the decay. In all cases, as the X-ray luminosity
increases, the characteristic frequencies shift to higher values.  The second
and third power-law indices of the double broken-power law model did not change
significantly during the outburst. To reduce the scatter in the frequency
relation, we fixed them to their average values, $\Gamma_2=1.37\pm0.05$ and
$\Gamma_3=1.66\pm0.07$. In contrast, the first power-law index, covering the
lower frequency range did show a smooth decline with luminosity (lower, right
panel in Fig.~\ref{freq}) and was let free during the fit. The QPO frequency
exhibits a very tight correlation regardless of the model used.  As in previous
studies \citep{revnivtsev09,doroshenko14}, we find that the QPO frequency is a
factor $\sim3-4$ lower than the break frequency.  For a detailed study of the
QPO variability during the 2020 outburst, we refer the reader to Ma et al.
(2022). 

 %---------------------------FIG 2-------------------------------------- 
%-----------------------------------------------------------------

\section{Discussion}

In this work, we have investigated the changes in the spectral and timing
continuum as a function of luminosity during the 2020 giant X-ray outburst of
\vtau. Before we discuss our results, let us summarize the different modes of
accretion in accreting pulsars.

The strong magnetic field  of the neutron star in accreting X-ray pulsars
disrupts the accretion flow at some distance from the neutron star surface and
forces the accreted matter to funnel down on the polar caps, creating hot spots
and an accretion column. The conditions prevailing in the accretion column
define two accetion regimes: super- and sub-critical. These two regimes differ
in the way the radiation pressure of the emitting plasma is capable of
decelerating the accretion flow. At high luminosities, in the super-critical
regime, the radiation pressure dominates and the braking of the accreting matter
flow is due to interaction with photons. A radiation-dominated shock stops the
flow at a certain distance from the neutron star surface
\citep{davidson73,basko76,lyubarskii82}. In the sub-critical regime, the
pressure of the radiation-dominated shock is not sufficient to stop the flow,
which continues its way to the neutron star surface where it is decelerated by 
multiple Coulomb scattering with thermal electrons and nuclear collisions with
atmospheric protons \citep{burnard91,harding94}.  A key parameter is the
luminosity at which the source transit from the sub- to the super-critical
regime. This is known as the critical luminosity, $L_{\rm crit}$
\citep{basko76,becker12,mushtukov15b}. 

At even lower luminosity, the Coulomb atmosphere dissipates and the matter goes
all the way to the neutron star surface, possibly passing through a gas-mediated
shock \citep{langer82}. The luminosity at which this transition occurs is
usually refer to as $L_{\rm coul}$ \citep[see eq.~(54) in][]{becker12}.
Collisions may also result in the excitation of electrons to upper Landau
levels, whose subsequently de-excitation generates cyclotron photons
\citep{mushtukov21}.  The energy spectrum at low accretion rates is
characterized by two broad components that peak at $\sim5-7$ keV and $\sim30-50$
keV \citep{tsygankov19}. In the model by \citet{mushtukov21}, the low energy
hump corresponds to Comptonized thermal radiation and the high energy hump to
Comptonized cyclotron photons.

Recent developments in the field has shown that this general picture should be
modified at the highest luminosity. In super-Eddington X-ray pulsars (so far
only Swift J0243.6+6124), a third regime has been suggested which would be
associated with the transition of the inner regions of the accretion disk from
the standard gas pressure dominated to the radiation pressure dominated state.
This transition would occur at a luminosity one order of magnitude higher than
the transition from the sub-critical to the super-critical regime
\citep{doroshenko20}. 

%---------------------------FIG 3-------------------------------------- 
%-----------------------------------------------------------------

\subsection{Hardness-intensity diagram and accretion regimes}

A model independent way to study the X-ray spectral continuum is through changes
in the count rate in two different energy bands (hardness ratio). The HID has
proven to be a useful tool to study spectral states not only in black hole
binaries and low-mass X-ray binaries \citep[see e.g.][]{vanderklis06,belloni10},
but also in HMXB \citep{reig08,reig13}.

\citet{reig13} investigated the X-ray color changes during type II outbursts of
nine BeXBs. They defined two main branches that were called horizontal and
diagonal and that they attributed to the sub- and super-critical accretion
states. Not all sources displayed the two branches. In fact, only four of the
nine studied sources transited to a super-critical state. \citet{reig13}
concluded that in order to see a transition to the super-critical state and
hence observed a clear diagonal branch in the HID the outburst peak source
luminosity must be several times the critical luminosity. 

Although the two branches can be clearly distinguished and identified by the 
direction of motion of the source in the HID --- in the horizontal branch the
source becomes harder as the intensity increases, whereas, in the diagonal
branch the source softens as it brightens --- this terminology is somehow
confusing as the horizontal branch normally appears inclined when a logarithmic
scale is used  \citep[confront e.g., Fig. 2 and 3 in][]{reig13}.

%---------------------------FIG 4-------------------------------------- 
\begin{figure}
\centering
\includegraphics[width=8cm]{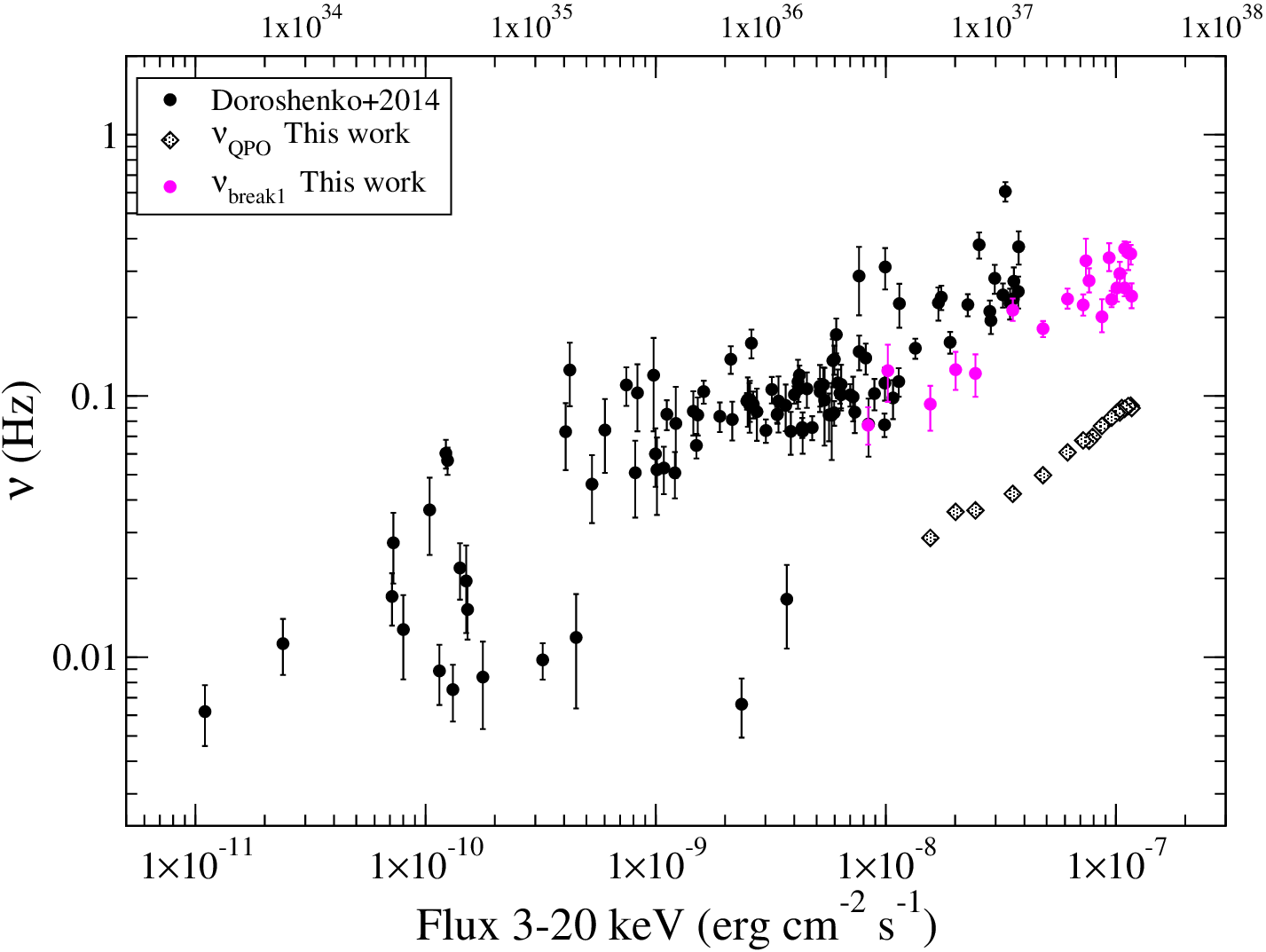}
\caption{Break frequencies as a function of X-ray flux. Data from {\it XMM}-Newton,
{\it RXTE}/PCA \citep[from][]{doroshenko14}, and {\it Insight-HXMT}/HE.}
\label{break-flux}
\end{figure}
%-----------------------------------------------------------------

In this work we shall refer to the branches by their dependency with intensity.
Therefore, the almost horizontal branch at the lowest count rate (green filled
circles in Fig.~\ref{hid}) which is formed by the first and last recorded
observations of the outburst  would be the low-intensity branch (LIB). The
inclined branch populated by all the observations of the rise and decay at
intermediate count rates would be intermediate-intensity branch (IIB). Finally,
the black circles that correspond to the peak of the outburst define the
high-intensity branch (HIB).  The horizontal and diagonal branches in the
terminology of \citet{reig13} would be the IIB and HIB, respectively. In the
case of \vtau, only the beginning of the HIB would be visible. The LIB (green
circles) is reported here for the first time.

Which and how many branches appear in the HID clearly depends on the sensitivity
of the detectors, the range in count rate covered by the observations and on the
value of the critical luminosity. The critical luminosity is different for
different sources as it strongly depends on the geometry of the accretion column
and the magnetic field \citep{basko76,becker12,mushtukov15b}. The larger the
magnetic field, the higher the critical luminosity and the longer time the
source spends in the sub-critical regime (IIB). Thus the reason that \vtau\
does not show a well developed super-critical branch (HIB) is the high value of
the critical luminosity in this system, $L_{\rm crit}\sim 6.7 \times 10^{37}$
\ergs\ 
\citep{kong21}. The maximum luminosity measured during the outburst is not
significantly larger than this value. Because the ratio $L_{\rm peak}/L_{\rm
crit}$ is not much larger than 1, the HIB (i.e. super-critical) branch does not
extend toward the left and only the initial instances of this branch are
observed in \vtau. Had the source luminosity increased further, we would have
presumably seen the HIB extending left toward lower hardness ratio as in 4U
0115+63, EXO 2030+375, KS 1947+300 and V 0332+53 \citep{reig13}. A transition to
the super-critical regime is supported by a detailed analysis of the energy
spectra, which shows sudden changes in the correlation of the photon index of
the power-law component \citep{mandal20} and  the cyclotron line \citep{kong21}
with luminosity.

The LIB (green circles in Fig.~\ref{hid}) is not present in any of the sources
(perhaps with the exception of XTE J0658-073) studied by \citet{reig13}.
Following the interpretation that the IIB and HIB correspond to the sub-critical
and super-critical accretion regimes, the LIB could be associated with the
transition from the Coulomb stopping to gas shock, that is, when both the
radiative shock and the Coulomb atmosphere have disappeared or are too weak. The
average luminosity during the LIB is $L_x \approx 6\times 10^{36}$ \ergs, which
approximately agrees with $L_{\rm coul}$ \citep{becker12} for typical parameters
of neutron stars and assuming a magnetic field of $\sim 4\times 10^{12}$ G. At
this and lower luminosity, the accretion flow is decelerated by collisions
between particles.  The LIB is characterized by a very fast softening of the
spectrum as the count rate decreases.  As the
accretion rate decreases, we would expect that the collisions of the accreting
particles with the electrons of the neutron star atmosphere would become less
efficient. Indeed, in the framework of the \citet{mushtukov21} model,  the
relative contribution of the thermal component with respect to the cyclotron
Comptonized component increases as the luminosity decreases \citep{tsygankov19}.

Another interesting finding is the sudden jump from the decaying track to the
rising track at around 700 c/s (2-30 keV) or $2\times 10^{37}$ \ergs\ in the low
energy hardness ratios (4-7/2-4 and 7-10/2-4) in Fig.~\ref{hid} (transition from
the magenta to the orange circles).  The effect is less important at higher
energy. The amplitude of change in harness ratio is 0.05 for HR=4-7/2-4, 0.02
for HR=7-10/3-5, 0.01 for HR=7-10/4-7, and non-existent for HR=15-30/10-15. It
is not clear what could lead to this effect. One may argue that the count rate
is not a good proxy for the accretion rate. However, the discontinuity remains
even when we use luminosity or flux instead of intensity. 

\subsection{Broad-band noise variability}

The accretion process generates strong aperiodic variability that manifest as
broad-band noise in the power spectrum of BeXBs
\citep{reig08,revnivtsev09,tsygankov12,reig13,doroshenko14,mushtukov19,doroshenko20}.
The PSDs display a characteristic break that has been
associated with the truncation radius of the accretion disk, at the location
where the accretion disk meets the magnetosphere \citep{revnivtsev09},  or
with the time scale of the dynamo process which is assumed to be responsible for
the initial fluctuations of viscosity \citep{mushtukov19} that propagate through
the disk and give rise to the aperiodic variability
\citep{lyubarskii97,churazov01}. 

QPOs are another prominent feature in the power spectrum of \vtau\ \citep[][see
also Ma et al. 2022]{finger96,camero12}. Most models locate the origin of the
QPO in the interplay between the accretion disk and the magnetosphere. The break
frequency would be associated with the time-scales on which accretion rate
fluctuations occur within the disk, while  QPOs would be associated with the
Keplerian time-scales at the inner edge of the accretion disk and the
magnetosphere boundary. The break frequency originating in a truncated disc is
expected to correlate with the Keplerian frequency at the magnetosphere.

A general property of the break and QPO frequencies is that they shift to higher
values as the X-ray luminosity increases. This results is naturally explained by
the relationship between mass accretion rate and the size of the magnetosphere. 
The magnetospheric radius scales with the mass accretion rate as $R_m \propto 
\dot{M}^{-2/7}$, as $\dot{M}$ increases, the luminosity increases and the
magnetosphere shrinks. The accretion disk radius decreases and the
characteristic frequency at the inner edge of the disk increases.
Figure~\ref{freq} shows the evolution of the characteristic frequencies of the
broad-band noise as a function of flux. 

The large peak luminosity displayed by \vtau\ during the 2020 outburst allows us
to study the frequency-luminosity correlation on a broader range in X-ray
luminosity as it is shown in  Fig~\ref{break-flux}. This correlation initially
covered two orders of magnitude in flux \citep{revnivtsev09}.
\citet{doroshenko14} extended it  at the lower end by analyzing observation
close to quiescence. Here we extend the correlation at the higher end up to
$10^{-7}$ erg cm$^{-2}$ s$^{-1}$. Thus, the correlation now holds for more than
four orders of magnitude in X-ray flux.

Although there is large scatter in the plot, we notice a flattening of the
correlation at the highest flux, already noticed in Fig.~\ref{freq}. In the
framework of the perturbation propagation model, the break frequency observed in
the PSD is related to the variability generated at the inner edge of the disk
\citep{revnivtsev09}. Therefore, it might be that the inner parts of the
accretion disk reached the closest possible distance to the neutron
star at the highest luminosities.

\subsection{Hysteresis}

Hysteresis patterns are commonly observed in BHB
\citep{miyamoto95,begelman14,weng21} and LMXB \citep{munoz-darias14}.   In
this context, hysteresis means that certain spectral and timing parameters have
different values depending on whether the source is in the rising phase of the
outburst or in the declining phase even though the luminosity is similar. 
Although the number of accreting pulsars investigated is not large, some cases
of hysteretic behavior have been reported. Hysteresis is not only seen in the
HID \citep{reig08}, but also in the pulse fraction \citep{wang20} and the spin
rate \citep{filippova17}.

In \vtau, the hysteresis pattern shows up, both in the spectral continuum and
the broad-band noise.  The color analysis reveals that for the same intensity,
the spectral hardness is larger during the rise than during the decay of the
outburst (compare the red and blue circles in Fig.~\ref{hid}). Also, the
hysteresis pattern is most prominent at low energies. The pattern disappears as
we consider higher energy bands ($E\simmore 4$ keV). The softer part of the
spectrum of accreting pulsars is associated with thermal emission from the polar
caps and/or base of the accretion column. In fact, \citet{kong21} found that two
blackbody components were needed to fit the spectrum of \vtau\ during the 2020
outburst. Interestingly, both the temperature and size of the emitting region of
the two thermal components exhibited hysteresis, with the temperatures being
larger  and the emitting region smaller during the rise \citep[see Fig.~5
in][]{kong21}.   The base of the accretion column and/or the polar cap area
appear to be more compact during its formation (rise) than during its
dissipation (decay). The different  path during the rise and decay is also
apparent in the photon index of the power-law component \citep{kong21}. 
Likewise,  the characteristic frequency of the main broad-band component
($\nu_{\rm Lor2}$ and $\nu_{\rm break1}$) was higher during the rise (red
circles in Fig.~\ref{freq}). The higher frequencies may simple
mean that the inner part of the accretion disk moved closer to the neutron star
during the rise compared to the same luminosity during the decay. 

The fact that the hysteresis is seen in both spectral and timing parameters set
tight constraints on the models that seek to explain this behavior. Spectral
variability occurs very close to the neutron star surface  in the accretion
column, while aperiodic variability originates in the accretion disk. The
structure that links these two emission sites is the magnetosphere. Therefore,
although there is no general consensus about the origin of the hysteresis effect
in accreting pulsars, it appears to be related to a different size of the
magnetosphere during the rise and decay, which would also translate into changes
in the configuration of the accretion column.  A smaller magnetospheric radius
was also invoked to explained the higher spin-up rate in the BeXB V\,0332+53
during the rise of its 2015 giant outburst \citep{doroshenko17}.

What causes this different size is unclear. It might be due to a variable
magnetic field strength \citep{cusumano16} or by a change in the emission region
geometry \citep[][but see \citealt{kylafis21}]{poutanen13,doroshenko17}. The
fact that the  energy of the cyclotron line was systematically smaller during
the rise \citep{kong21} might indicate a weaker magnetic field.
However, it is not clear how the magnetic field strength can change on such time
scales. In V\,0332+53, the evolution of the energy of the cyclotron line was
opposite to that of \vtau, with a drop of the cyclotron line energy during the
declining part. Therefore, it is difficult to think of a mechanism that would
change the magnetic field strength in opposite ways for two apparently similar
sources. Equally, it is not clear what could lead to the different evolution of
the inner disk radius in the two sources that would favour a smaller
magnetosphere during the rise in one case and a larger magnetosphere in the
other source, also during the rising phase. As pointed out by \citet{kong21},
perhaps the different spin period of the two sources, V\,0332+53 is a fast
pulsar ($P_{\rm spin}=4.4$ s), while \vtau\ is a slow pulsar ($P_{\rm spin}=103$
s) leads to a different interplay between the accretion column and the 
magnetosphere. We also note that the optical counterpart to V\,0332+53 is more
massive (O8-9V star) and orbits at a significantly closer distance ($P_{\rm
orb}=34$ days) than that of \vtau\ (B0III star, $P_{\rm orb}=111$ days).

\section{Conclusion}

The 2020 November bright X-ray outburst of \vtau\ allowed us to study the
spectral-timing properties of this accreting pulsar at very high luminosity.
Despite the different origin of the spectral variability (accretion column) and
the broad-band noise (accretion disk), both kind of variability provide evidence
for a different behavior depending on whether the source was on the rise or on
the decay of the outburst. The hysteresis pattern is dominant at intermediate
luminosity and low energy and appears to be related to a variable magnetospheric
radius. At similar luminosity, the inner parts of the accretion disk would move
closer to the neutron star during the rise. The hardness-intensity diagram
appears as a useful tool to identify accretion regimes. We observe three
different branches that we identified with accretion at $L_X \simmore L_{\rm
crit}$ (HIB) and two branches associated with the sub-critical regime when $
L_{\rm coul} < L_X < L_{\rm crit}$ (IIB) and when $L_X\sim L_{\rm coul}$ (LIB).
Because of its high magnetic field, the critical luminosity in \vtau\ is high
and the source only traces the beginning of the HIB.  We have extended the
correlation between the break frequency of the broad-band noise and the
luminosity above $L_X > 2 \times 10^{37}$ \ergs\ and now it holds over four
orders of magnitude in luminosity.

\begin{acknowledgements}

ZS and TL acknowledge the supports of the National Key R\&D Program of China
(2021YFA0718500) and the National Natural Science Foundation of China under
grant U1838201, U1838202, and 11733009. L.T. acknowledges funding support from
the National Natural Science Foundation of China (NSFC) under grant No. 12122306
and the CAS Pioneer Hundred Talent Program Y8291130K2. This work has made use of
the data from the Insight-HXMT mission, a project funded by China National Space
Administration (CNSA) and the Chinese Academy of Sciences (CAS).

\end{acknowledgements}

\bibliographystyle{aa}
\bibliography{../../../artBex_bib.bib} % if your bibtex file is called bhb.bib

\end{document}